\documentclass[12pt,epsf]{article}
\usepackage{graphicx,epsfig,amsmath,amssymb}
\begin{document}

\def\a{\alpha}
\def\b{\beta}
\def\c{\varepsilon}
\def\d{\delta}
\def\e{\epsilon}
\def\f{\phi}
\def\g{\gamma}
\def\h{\theta}
\def\k{\kappa}
\def\l{\lambda}
\def\m{\mu}
\def\n{\nu}
\def\p{\psi}
\def\q{\partial}
\def\r{\rho}
\def\s{\sigma}
\def\t{\tau}
\def\u{\upsilon}
\def\v{\varphi}
\def\w{\omega}
\def\x{\xi}
\def\y{\eta}
\def\z{\zeta}
\def\D{\Delta}
\def\G{\Gamma}
\def\H{\Theta}
\def\L{\Lambda}
\def\F{\Phi}
\def\P{\Psi}
\def\S{\Sigma}

\def\o{\over}
\def\beq{\begin{eqnarray}}
\def\eeq{\end{eqnarray}}
\newcommand{\gsim}{ \mathop{}_{\textstyle \sim}^{\textstyle >} }
\newcommand{\lsim}{ \mathop{}_{\textstyle \sim}^{\textstyle <} }
\newcommand{\vev}[1]{ \left\langle {#1} \right\rangle }
\newcommand{\bra}[1]{ \langle {#1} | }
\newcommand{\ket}[1]{ | {#1} \rangle }
\newcommand{\EV}{ {\rm eV} }
\newcommand{\KEV}{ {\rm keV} }
\newcommand{\MEV}{ {\rm MeV} }
\newcommand{\GEV}{ {\rm GeV} }
\newcommand{\TEV}{ {\rm TeV} }
\def\diag{\mathop{\rm diag}\nolimits}
\def\Spin{\mathop{\rm Spin}}
\def\SO{\mathop{\rm SO}}
\def\O{\mathop{\rm O}}
\def\SU{\mathop{\rm SU}}
\def\U{\mathop{\rm U}}
\def\Sp{\mathop{\rm Sp}}
\def\SL{\mathop{\rm SL}}
\def\tr{\mathop{\rm tr}}

\def\IJMP{Int.~J.~Mod.~Phys. }
\def\MPL{Mod.~Phys.~Lett. }
\def\NP{Nucl.~Phys. }
\def\PL{Phys.~Lett. }
\def\PR{Phys.~Rev. }
\def\PRL{Phys.~Rev.~Lett. }
\def\PTP{Prog.~Theor.~Phys. }
\def\ZP{Z.~Phys. }

\newcommand{\bea}{\begin{eqnarray}}
\newcommand{\eea}{\end{eqnarray}}
\newcommand{\bear}{\begin{array}}
\newcommand {\eear}{\end{array}}
\newcommand{\bef}{\begin{figure}}
\newcommand {\eef}{\end{figure}}
\newcommand{\bec}{\begin{center}}
\newcommand {\eec}{\end{center}}
\newcommand{\non}{\nonumber}
\newcommand {\eqn}[1]{\beq {#1}\eeq}
\newcommand{\la}{\left\langle}
\newcommand{\ra}{\right\rangle}
\newcommand{\ds}{\displaystyle}
\newcommand{\hyp}{\,{\rm \mathchar`-}\,}
\def\SEC#1{Sec.~\ref{#1}}
\def\FIG#1{Fig.~\ref{#1}}
\def\EQ#1{Eq.~(\ref{#1})}
\def\EQS#1{Eqs.~(\ref{#1})}
\def\TEV#1{10^{#1}{\rm\,TeV}}
\def\GEV#1{10^{#1}{\rm\,GeV}}
\def\MEV#1{10^{#1}{\rm\,MeV}}
\def\KEV#1{10^{#1}{\rm\,keV}}
\def\lrf#1#2{ \left(\frac{#1}{#2}\right)}
\def\lrfp#1#2#3{ \left(\frac{#1}{#2} \right)^{#3}}
\def\oten#1{ {\mathcal O}(10^{#1})}


\baselineskip 0.7cm

\begin{titlepage}

\begin{flushright}
TU-921
\end{flushright}

\vskip 1.35cm
\begin{center}
{\large \bf
    A Gravitino-rich Universe
    }
\vskip 1.2cm

\renewcommand{\thefootnote}{\fnsymbol{footnote}}

    Kwang Sik Jeong$^\ast$\footnote[0]{$^\ast$ email: ksjeong@tuhep.phys.tohoku.ac.jp}
    and
    Fuminobu Takahashi$^\dagger$\footnote[0]{$^\dagger$ email: fumi@tuhep.phys.tohoku.ac.jp}

\vskip 0.4cm

{\it Department of Physics, Tohoku University, Sendai 980-8578, Japan}\\

\vskip 1.5cm

\abstract{
The gravitino may well play an important role in cosmology, not only because its interactions
are Planck-suppressed and therefore long-lived, but also because it is copiously produced
via various processes such as particle scatterings in thermal plasma, and (pseudo) modulus
and inflaton decays.
We study a possibility that the early Universe was gravitino-rich from various aspects.
In particular, a viable cosmology is possible,
if high-scale supersymmetry is realized in nature as suggested by the recent discovery of
the  standard-model like  Higgs boson of mass about $125 \,{\rm \mathchar`-}\, 126$\,GeV.
We find that  the Universe can be even gravitino-dominated, in which case
there will be an entropy dilution by the gravitino decay. If the gravitino abundance is proportional
to the reheating temperature,  both the maximal baryon asymmetry in leptogenesis
and the dark matter from the gravitino decay become independent of the reheating temperature.
The dark matter candidate is the Wino-like neutralino, whose mass is
suppressed compared to the anomaly-mediation relation.
}
\end{center}

\end{titlepage}

\setcounter{page}{2}

\section{Introduction}
\label{sec:1}
The recent discovery of the standard-model (SM) like Higgs boson of mass about
$125$  $\,{\rm \mathchar`-}\,$ $126$\,GeV~\cite{:2012gk,:2012gu}
may imply high-scale supersymmetry (SUSY)~\cite{Okada:1990gg,Giudice:2011cg}.
SUSY relates fermions to bosons and vice versa,
and adds the superpartner of each  particle in nature; hence, in a supersymmetric theory
there will be a fermion that partners with the graviton, called the gravitino, which
becomes massive via the super-Higgs mechanism after SUSY gets spontaneously broken.
The interactions of the gravitino are suppressed by either the Planck scale or the SUSY breaking
scale, and therefore it is naturally long-lived.

The gravitinos are produced in various processes. In the early Universe, the gravitinos are produced by
particle scatterings in thermal plasma~\cite{Bolz:2000fu,Pradler:2006qh,Rychkov:2007uq}.
Also, it is known that any scalar field with a non-vanishing vacuum
expectation value (VEV) generically decays into gravitinos, if kinematically allowed.
In particular, a large amount of gravitinos can
be produced by the (pseudo) modulus~\cite{moduli,Dine:2006ii,Endo:2006tf} and
inflaton~\cite{Kawasaki:2006gs,Asaka:2006bv,Endo:2006qk,Endo:2007ih,Endo:2007sz} decays.

If the gravitinos are abundant in the early Universe, they may play an important
role in cosmology, because of their longevity.  Indeed, it is well-known that too
many gravitinos significantly affect the evolution of the Universe
in contradiction with observations;
{\it the cosmological gravitino problem}~\cite{Weinberg:1982zq,Ellis:1982yb,Krauss:1983ik}.
For instance, if the gravitino mass is around the weak scale, it typically decays during
big bang nucleosynthesis (BBN), altering the light element abundances in contradiction
with observations.
In the case of high-scale SUSY, however, the gravitino can decay before BBN,
and the success of the standard BBN remains intact. Thus, the high-scale SUSY may lead to
a viable cosmology even if many gravitinos are produced in the early Universe.
The main purpose of this paper is to
study such {\it  gravitino-rich Universe} from both cosmological and phenomenological aspects.

What is peculiar to the gravitino is that
it is coupled to {\it any} sectors in nature. Thus, once produced, the gravitinos will
distribute their energy to all the lighter particles including the SM particles.
For instance, the gravitino decay produces the lightest ordinary supersymmetric particle.
If it is the lightest supersymmetric particle (LSP) and if the R-parity is conserved,
the LSPs thus produced may
account for the observed dark matter (DM) abundance.
Alternatively,  if there are light degrees of freedom
in a hidden sector, the gravitino necessarily decays into them.
Some of the hidden-sector particles thus produced
may contribute to DM, if they are long-lived and non-relativistic. If they are
relativistic at the BBN and recombination epoch, they will increase the expansion rate, which
may account for the dark radiation hinted by recent observations~\cite{Komatsu:2010fb,Dunkley:2010ge}.
(See e.g. Refs.~\cite{Ichikawa:2007jv, Jaeckel:2008fi, Nakayama:2010vs, Kobayashi:2011hp, Jeong:2012hp, Choi:2012zn,Cicoli:2012aq,Higaki:2012ar}  for the
models.)
Also, if the inflation took place in a hidden sector,
the only way to reheat the visible sector may be through the gravitino production;
the gravitino may play a role of messenger because of its universal couplings.
Thus, the gravitino-rich Universe seems to have various interesting implications.

\vspace{5mm}
In this paper we shall study if the Universe becomes gravitino-rich by considering
several production processes of the gravitinos; thermal production as well as
non-thermal production by the (pseudo) modulus and inflaton decays.
We find that the gravitino-rich Universe is realized for a wide range of parameters,
and the Universe even becomes gravitino-dominated in some cases.
In the latter case, there is an entropy dilution by the gravitino decay. Interestingly,
the gravitino abundance and therefore its dilution factor are proportional to the
reheating temperature in the case of thermal production and
the non-thermal production by the pseudo-modulus decay. Then,
the maximal baryon asymmetry in thermal leptogenesis becomes independent of the reheating
temperature. Assuming the
Wino-like LSP of mass several hundred GeV, the LSP abundance can account for
the observed DM. For this, the anomaly mediation relation for the gaugino masses must
be modified, requiring a mild cancellation of  order $10$\% in the Wino mass.
Note that both the baryon asymmetry and the neutralino LSP abundance are independent
of the reheating temperature in this case.

\vspace{5mm}
The rest of this paper is organized as follows. In Sec.~\ref{sec:2}
we briefly discuss the gravitino lifetime and abundance.
We study various production processes of the gravitinos in Sec.~\ref{sec:3}.
In Sec.~\ref{sec:4} we discuss cosmological aspects
of the gravitino-rich Universe, and study the maximal baryon
asymmetry in the thermal leptogenesis as well as the DM abundance.
In Sec.~\ref{sec:5} we discuss how the gaugino mass relation can be deviated from
the anomaly mediation relation. The last section is devoted for discussion and conclusions.

\section{Gravitino abundance}
\label{sec:2}

Let us summarize here the abundance and lifetime of the gravitino.
For the moment we assume the minimal particle content, namely, the minimal supersymmetric
standard model (MSSM) particles and the gravitino.
Later we shall consider a case that there are light degrees of freedom in the hidden sector.

The decay rate of the gravitino into the MSSM particles is approximately given by
\beq
\Gamma_{3/2}(\psi_{3/2} \rightarrow {\rm\,MSSM}) \;\simeq\; \frac{193}{384 \pi} \frac{m_{3/2}^3}{M_P^2},
\eeq
where $m_{3/2}$ is the gravitino mass, $M_P \simeq 2.4 \times 10^{18}$\,GeV is the reduced Planck mass,
and we have assumed that the gravitino is heavier than all the MSSM particles, for simplicity.
The decay temperature of the gravitino, $T_{3/2}$, is defined by
\bea
T_{3/2} &\equiv& \lrfp{\pi^2 g_*(T_{3/2})}{90}{-\frac{1}{4}} \sqrt{\Gamma_{3/2} M_P},\\
       &\simeq& 0.15 {\rm \,GeV} \lrfp{g_*(T_{3/2})}{80}{-\frac{1}{4}} \lrfp{m_{3/2}}{10^3{\rm\,TeV}}{\frac{3}{2}},
       \label{decayT}
\eea
where $g_*(T)$ counts the relativistic degrees of freedom at temperature $T$.
If the gravitino is heavier than several tens TeV, its decay temperature is above
a few MeV, avoiding the tight BBN constraint~\cite{LowTRH}.
We adopt $m_{3/2} = 10^3$\,TeV as a reference value
in the following.

Let us quantify the gravitino abundance in terms of the ratio of the gravitino number density to the
entropy density;
\beq
Y_{3/2}\;\equiv\; \frac{n_{3/2}}{s_i},
\eeq
where $s_i$ represents the entropy density originated from the inflaton decay. Note that $s_i$  does
not include the entropy produced by the gravitino decay.

If the gravitinos are abundant, they may come to dominate the energy density of the Universe,
since the gravitino is massive and long-lived.
The gravitino-dominated Universe is realized if
\beq
Y_{3/2}\;>\; \frac{3}{4} \frac{T_{3/2}}{m_{3/2}}.
\eeq
This is equivalent to the condition that the entropy produced by the gravitino, $s_{3/2}$,
be larger than the pre-existing entropy, $s_i$.
Using \EQ{decayT}, this condition can be rewritten as
\beq
Y_{3/2} \;\gtrsim\; 1.5 \times 10^{-7}  \lrfp{g_*(T_{3/2})}{80}{-\frac{1}{4}}
\lrfp{m_{3/2}}{10^3{\rm\,TeV}}{\frac{1}{2}}.
\label{gra_dom}
\eeq
If this condition is met, the entropy dilution factor $\Delta$ is given by
\bea
\Delta \;\equiv\; \frac{s_{3/2}+s_i}{s_i} &\simeq&  \frac{n_{3/2}}{s_i} \lrfp{n_{3/2}}{s_{3/2}}{-1},\\
&\simeq& 90  \lrfp{g_*(T_{3/2})}{80}{\frac{1}{4}}
\lrfp{m_{3/2}}{10^3{\rm\,TeV}}{-\frac{1}{2}} \lrf{Y_{3/2}}{10^{-5}},
\label{eq:ed}
\eea
where $s_{3/2}$ represents the entropy density produced by the gravitino decay, and
we have used $(n_{3/2}/s_{3/2}) = 3T_{3/2}/4m_{3/2}$.
Noting that $Y_{3/2} \sim 10^{-2}$ when gravitinos are in thermal equilibrium,
one can see that the dilution factor varies from $1$ to $\sim 10^5$ for the gravitino
mass around $\TEV{3}$.

It is possible that a modulus field dominates the Universe after the reheating, and decays into gravitinos.
If the branching fraction of the modulus decay into gravitinos is large enough, both the modulus and gravitino
decays produce entropy.  In this case the total entropy dilution factor is the same as \EQ{eq:ed},
and it can be also expressed in terms of the modulus abundance as
\bea
\Delta  &\simeq&  \frac{\rho_X}{s_i} \frac{n_{3/2}}{\rho_X} \lrfp{n_{3/2}}{s_{3/2}}{-1},\\
&\simeq& \frac{4}{3} B_{3/2} N_{3/2} \frac{m_{3/2}}{m_X T_{3/2}} \lrf{\rho_X}{s_i},
\label{eq:ed2}
\eea
where $B_{3/2}$ denotes the branching fraction into gravitinos, $N_{3/2}$ is the average
number of gravitinos for the decay modes into gravitinos, $m_X$ and $\rho_X$ are
the mass and energy density of the modulus $X$. When the modulus mainly decays into
a pair of gravitinos, we expect $B_{3/2} \simeq 1$ and $N_{3/2} = 2$.

If (\ref{gra_dom}) is not satisfied, the gravitinos do not dominate the energy density of the Universe,
but they can still have a significant impact on cosmology. For instance,
 if its abundance is greater than $\sim 10^{-12}$, the LSPs produced by the gravitino
decay tend to overclose the Universe, as long as the annihilation of the LSPs is not effective and
the LSP mass is larger than or comparable to ${\cal O}(100)$\,GeV. We
therefore adopt the criterion for the gravitino-rich Universe as,
\setlength{\fboxrule}{1pt}
\beq
\boxed{
\hspace{3mm}
{\rm Gravitino\mathchar`-rich~ Universe~}: ~~Y_{3/2} \;>\; 10^{-12}
}
\label{grU}
\eeq
Of course, there are various possibilities, for instance, the R-parity may be violated, and the QCD axion
may be the dominant component of DM, or there may be other lighter SUSY particles in the hidden sector, and so on.
A different criterion should be adopted depending on the scenarios of interest. To be concrete, however, we adopt
the above criterion in this paper.

\section{Production processes}
\label{sec:3}

Now we consider various processes for the gravitino production.

\subsection{Thermal production}
Gravitinos are produced by particle scatterings in thermal plasma,
and its abundance is given by~\cite{Bolz:2000fu,Pradler:2006qh,Rychkov:2007uq}
\begin{equation}
    Y_{3/2}^{(\rm TP)}\; \simeq\;  2\times 10^{-12} 
    \left( \frac{T_R}{10^{10}\,{\rm GeV}} \right),
    \label{y32th}
\end{equation}
where we have omitted the logarithmic dependence on $T_R$ as well as terms that depend on the
gaugino masses.  The definition of $T_R$ is given by
\beq
T_R \;\equiv\; \lrfp{\pi^2 g_{\rm *}(T_R)}{90}{-\frac{1}{4}} \sqrt{\Gamma_\phi M_P},
\label{TRdef}
\eeq
where $\Gamma_\phi$ denotes the total decay rate of the inflaton, and we assume that the inflaton
mainly decays into the MSSM sector. Thus, the gravitino abundance increases in proportion to
the reheating temperature.

The gravitino-rich Universe is realized if the reheating temperature satisfies
\beq
T_R \;\gtrsim\;  \GEV{10},
\eeq
while the gravitino dominates the Universe if the reheating
temperature is
\beq
T_R \;\gtrsim\; \GEV{15}\,\lrfp{g_*(T_R)}{80}{-\frac{1}{4}} \lrfp{m_{3/2}}{10^3{\rm\,TeV}}{\frac{1}{2}}.
\eeq
Thus, an extremely high reheating temperature is required for the gravitino to dominate the Universe.

Note however that the expression for the gravitino abundance (\ref{y32th}) does not take account of
the longitudinal mode. The contribution from the longitudinal mode, namely the goldstino, becomes
relevant if the gauginos are heavier than the gravitino. Furthermore,
if the inflaton decays into the SUSY breaking sector,\footnote{
The inflaton itself may be a part of the SUSY breaking sector.
}
the would-be goldstino may be thermalized,
leading to $Y_{3/2}^{(\rm TP)} \sim 10^{-2}$, where we have assumed that the relativistic degrees
of freedom is of order $10^2$. Thus, it should be kept in mind that the above estimate on
the gravitino abundance
(\ref{y32th}) depends on the assumption about the reheating process and the mass spectrum.

\subsection{Non-thermal production}
Now let us consider non-thermal production of gravitinos. In the following we consider
a pseudo modulus in the SUSY breaking sector, a general modulus field, and an inflaton as possible
 source of gravitinos.

\subsubsection{Pseudo modulus field in the SUSY breaking sector}
We first consider a pseudo-modulus decay into gravitinos.
A pseudo modulus in the SUSY breaking
sector is a plausible candidate for the gravitino production
 because of the following reasons; (i) the existence of the pseudo modulus is generic
in the SUSY breaking models; (ii) the coherent oscillations are produced because the potential minimum
for the pseudo modulus during inflation is generically deviated from the low-energy minimum; (iii)
it predominantly decays into a pair of gravitinos.

We study the low-energy effective theory of O'Raifeartaigh type SUSY
breaking model.  After integrating out the massive fields, the K\"ahler and super-potentials
are
\bea
\label{kp}
K &=& |z|^2  -\frac{|z|^4}{\Lambda^2} + \cdots,\\
W &=& \mu^2 z  + W_0,
    \label{spp}
\eea
where $z$ is a pseudo modulus field, $\Lambda$ is a cut-off scale, $\mu$ represents the SUSY breaking scale,
and the constant $W_0 \simeq m_{3/2} M_P^2$  is fixed so that the cosmological constant almost vanishes
at present.
We can assign an R-charge $2$ on $z$, which is explicitly broken by $W_0$ down to $Z_{2R}$.
All the parameters are set to be real by an appropriate redefinition of $z$ and the U(1)$_R$ transformation.

The $z$ is stabilized at the potential minimum,
\begin{equation}
    \langle z\rangle = \frac{2\sqrt{3} m_{3/2}^2 M_P}{m_z^2},  \label{SVEV}
\end{equation}
where $m_z$ is the mass of $z$ given by
\beq
m_z \;=\; \frac{2 \mu^2}{\Lambda}.
\eeq
The F-term of $z$ is given by $F_z \simeq - \mu^2 \simeq \sqrt{3} m_{3/2} M_P$, and SUSY is indeed
broken.
The precise value of $\Lambda$ depends on details of the SUSY breaking. We simply assume here that
$\Lambda$ is much smaller than the Planck scale so that $z$ is heavier than the gravitino.

\begin{figure}[t]
\begin{center}
\begin{minipage}{16.4cm}
\centerline{
{\hspace*{0cm}\epsfig{figure=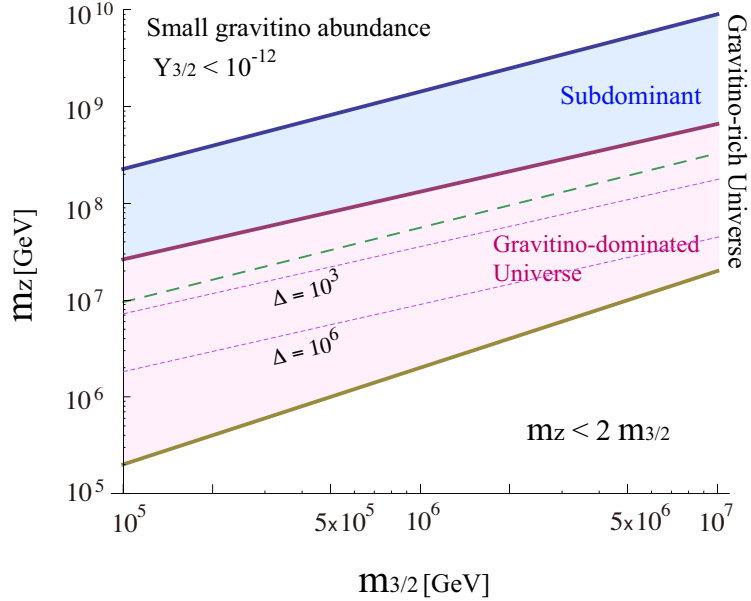,angle=0,width=10cm}}
}
\caption{
    In the shaded (red and blue) regions, the Universe becomes gravitino-rich (see Eq.~(\ref{grU})).
    The gravitino dominates the Universe in the lower shaded (red) region.
    The pseudo modulus also dominates the Universe in the region below the dashed line.
    We have fixed $T_R = \GEV{9}$, but the lines are only weakly dependent on $T_R$.
    See the text for details.
}
\label{fig:pseudo modulus}
\end{minipage}
\end{center}
\end{figure}

Suppose that the U(1)$_R$ symmetry remains a good symmetry during inflation and that
the inflation scale, $H_{\rm inf}$, is larger than $m_z$. Then, if $z$ acquires a positive Hubble-induced
mass, $z$ is stabilized near the
origin during inflation, and starts to oscillate with an amplitude of $\la z \ra$
when $H \sim m_z$ after inflation.\footnote{
Even if the Hubble parameter during inflation exceeds the dynamical scale, it is possible
that $z$ is stabilized near the origin,  when the Hubble parameter becomes so small after inflation
that the low-energy description (\ref{kp}) and (\ref{spp}) are valid.
} Thus the pseudo-modulus abundance is given by
\begin{equation}
    \frac{\rho_z}{s_i} \;\simeq\; 3T_{R}\left(\frac{m_{3/2}}{m_z}\right)^4,
    \label{pm_ab}
\end{equation}
where we have assumed that the pseudo modulus starts to oscillate before the reheating.
On the other hand, if the Hubble parameter during inflation is smaller than $m_z$,
the pseudo-modulus abundance gets suppressed.
If the inflaton at the potential minimum is heavier than $m_z$,  the pseudo modulus cannot
 follow the change of the potential minimum, and the coherent oscillations are induced~\cite{Nakayama:2011wqa,Higaki:2012ba}.
 The abundance is given by
\begin{equation}
    \frac{\rho_z}{s_i} \;\simeq\; 3T_{R}\left(\frac{m_{3/2}}{m_z}\right)^4 \lrfp{H_{\rm inf}}{m_z}{4}.
\end{equation}
If the inflaton mass at the potential minimum is lighter than $m_z$, no coherent oscillations are
induced. In the following analysis we use the pseudo-modulus abundance (\ref{pm_ab}).
As one can see from Fig.~\ref{fig:pseudo modulus}, this is the case for the inflation scale larger than
$10^9 \sim 10^{10}$\,GeV, which covers many inflation models.

The quartic coupling in the K\"ahler potential induces the $z$ decay into the goldstino pair
with the decay rate
\begin{equation}
    \Gamma(z\to \tilde z\tilde z) \;\simeq\; \frac{1}{96\pi}\frac{m_z^5}{m_{3/2}^2M_P^2}.
    \label{StoGG}
\end{equation}
As long as $z$ does not have any sizable couplings with the SM sector,\footnote{
Planck-suppressed couplings do not change the argument.
}
it predominantly decays into a pair of gravitinos. Thus, the gravitino abundance from
the pseudo-modulus decay is given by
\beq
Y_{3/2}^{(z)}\;\simeq\; 6 \times 10^{-7} \lrf{T_R}{\GEV{9}} \lrfp{m_{3/2}}{\TEV{3}}{4}
\lrfp{m_z}{\GEV{8}}{-5}.
\eeq
The Universe becomes gravitino-rich if $Y_{3/2}^{(z)}$ is greater than $10^{-12}$, namely,
\beq
m_z \;\lesssim\; 1\times 10^{9}\,{\rm GeV}\,
\lrfp{T_R}{\GEV{9}}{\frac{1}{5}} \lrfp{m_{3/2}}{\TEV{3}}{\frac{4}{5}}.
\label{pseudo modulus2}
\eeq
The gravitino-dominated Universe is realized if
\beq
m_z \;\lesssim\; 1\times 10^{8}\,{\rm GeV}\,\lrfp{g_*}{80}{\frac{1}{20}} \lrfp{T_R}{\GEV{9}}{\frac{1}{5}} \lrfp{m_{3/2}}{\TEV{3}}{\frac{7}{10}}.
\label{pseudo modulus1}
\eeq
In Fig.~\ref{fig:pseudo modulus}, the Universe becomes gravitino-rich in the shaded (red and blue) regions.
In the lower shaded (red) region, the gravitino dominates the Universe, while it does not in the upper shaded (blue)
region. The dotted (purple) lines show the contours of the entropy dilution factor, $\Delta = 10^3$ and $10^6$.
In the upper left region, the condition (\ref{pseudo modulus2}) is not satisfied
and the gravitino abundance is small,
while the decay into a pair of gravitinos is kinematically forbidden in the lower right region.
In the region below the dashed line, the pseudo modulus dominates the Universe before the decay.
We have fixed $T_R = \GEV{9}$ in the figure,
but the conditions on $m_z$ are only weakly dependent on $T_R$; both the upper and middle solid lines (the dashed line)
shift in proportion to $\propto T_R^{\frac{1}{5}} \,(T_R^{2/13})$.

\subsubsection{Modulus field}
Next we consider a modulus field as a source of gravitinos.
In superstring theories, moduli fields necessarily appear at low energies through
compactifications. Most of these moduli must be stabilized  in order to get a sensible low-energy theory,
since the moduli determine all the physically relevant quantities such as the
size of the extra dimensions, physical coupling constants, and even the SUSY breaking
scale. Many moduli fields are stabilized by flux compactifications~\cite{Grana:2005jc, Blumenhagen:2006ci},
and the remaining ones can be stabilized a la KKLT~ \cite{Kachru:2003aw}.

The detailed properties of the modulus depend on the compact geometry, brane configurations,
and stabilization mechanism.
Here we consider a modulus field $X$  stabilized a la KKLT; it has a mass  heavier
than the gravitino, and its F-term is suppressed by the mass as $F_X \sim m_{3/2} \la X \ra / m_X$, where
$m_X$ denotes the mass of $X$.
During inflation the potential of $X$ is considered to be deformed.\footnote{
There is an upper bound on the inflation scale in order to avoid the destabilization and run-away. Even for
the inflation satisfying the bound, the cosmological moduli problem still exists~\cite{Higaki:2012ba}.
} Then the potential minimum may be deviated from the low-energy minimum, leading to coherent oscillations
after inflation. We assume that the energy stored in the coherent oscillations of $X$ dominates the Universe.
The entropy density $s_i$ should be interpreted as that from the modulus decay in this subsection.

We also assume that the modulus $X$ has
Planck-suppressed couplings to the visible sector so that its decay rate can be expressed as
\beq
\Gamma_X \;=\; \frac{c}{4 \pi} \frac{m_X^3}{M_P^2},
\eeq
where $c$ is a numerical constant of order unity. The decay temperature of the modulus is
\bea
T_X &=& \lrfp{\pi^2 g_*(T_X)}{90}{-\frac{1}{4}} \sqrt{\Gamma_X M_P},\\
&\simeq& 3{\rm\,GeV} \lrfp{g_*(T_X)}{80}{-\frac{1}{4}} \lrfp{m_X}{\GEV{7}}{\frac{3}{2}}.
\eeq

It was shown in Refs.~\cite{moduli,Dine:2006ii,Endo:2006tf} that such a modulus generically
decays into a pair of gravitinos
and the branching fraction  is sizable: $B_{3/2} = {\cal O}(0.01 \,{\rm \mathchar`-} \,0.1)$.\footnote{
Precisely speaking, this is the case if the SUSY breaking field is heavier than the modulus mass, or if there is a
SUSY breaking field that is singlet under any symmetries.}
Then the gravitino abundance from the modulus decay is estimated as
\bea
Y_{3/2}^{(X)} &=& 2 B_{3/2} \frac{3}{4} \frac{T_X}{m_X},\\
&\simeq&  6  \times 10^{-7} B_{3/2} \sqrt{c} \lrfp{g_*}{80}{-\frac{1}{4}} \lrfp{m_X}{\GEV{7}}{\frac{1}{2}}.
\label{y32x}
\eea
Thus, the Universe becomes gravitino-rich for a wide range of the modulus mass and the branching
fraction of the gravitino production. In particular,  if the
branching fraction into gravitinos is of order $0.1$, the gravitino even dominates the Universe.
(Compare (\ref{y32x}) with (\ref{gra_dom}).)

\subsubsection{Inflaton}
In a series of works~\cite{Dine:2006ii,Endo:2006tf,Kawasaki:2006gs,Asaka:2006bv,
Endo:2006qk,Endo:2007ih,Endo:2007sz,Endo:2007cu}, it was revealed that the inflaton generally
decays into gravitinos. The gravitino production rate depends on whether there is a SUSY breaking
field that is singlet under any symmetries. Such a singlet is required in the gravity mediation to generate
gaugino masses of the correct size, while it is not necessary in the gauge and anomaly mediation.
We assume that there is no such a singlet for the moment.

The gravitinos are produced by various processes;
(a) the gravitino pair production~\cite{Dine:2006ii,Endo:2006tf,Kawasaki:2006gs,Asaka:2006bv};
(b) decay into the SUSY breaking sector at tree level~\cite{Endo:2006qk}; (c) anomaly-induced decay into
the SUSY breaking sector at one-loop level~\cite{Endo:2007ih}.
The gravitino production rate can be expressed as
\beq
\Gamma_{3/2} = \frac{x}{32 \pi} \lrfp{\la \phi \ra}{M_P}{2} \frac{m_\phi^3}{M_P^2},
\eeq
where $m_\phi$ is the inflaton mass, and $\la \phi \ra$ a VEV of the inflaton.  Here it should be noted that
$\la \phi \ra$ is evaluated at the potential minimum after inflation.
The precise value of the numerical coefficient $x$ depends on the
production processes, possible non-renormalizable couplings in the
K\"ahler potential, and the details of the SUSY breaking models~\cite{Endo:2007sz}.  To be concrete, let us
assume the minimal K\"ahler potential and the dynamical SUSY breaking
(DSB) with a dynamical scale $\Lambda$. As we have seen before, the SUSY breaking field $z$ can acquire a mass
$m_z$ heavier than $m_{3/2}$,  and we assume
$m_z\sim\Lambda \sim \sqrt{m_{3/2} M_P}$ in the following.\footnote{The gravitino production can be
suppressed if the SUSY breaking field is lighter than the inflaton and the inflaton mass is
below the dynamical scale~\cite{Endo:2007cu,Nakayama:2012hy}. }

 For a low-scale inflation model with $m_\phi < \Lambda$, the process (a) becomes
effective, and $x = 1$.  On the other hand, for the inflaton mass
larger than $\Lambda$, the processes (b) and (c) become effective
instead. The inflaton decays into the hidden quarks in the SUSY
breaking sector via Yukawa couplings of the hidden particles (process (b)), or into the hidden
gauge sector via anomalies (process (c)).  Since the hidden quarks and
gauge bosons (and gauginos) are energetic when they are produced, they
will form jets and produce hidden hadrons through the
strong gauge interactions.  The gravitinos are likely produced by the
decays of the hidden hadrons.
We denote the averaged number of the gravitinos produced per each jet as $N_{3/2}$.
Then $x$ is given by~\cite{Endo:2007sz}
\beq
x \;\simeq\; \frac{N_{3/2}}{8 \pi^2} \left(\frac{1}{2} N_y |Y_h^2| +
N_g \alpha_h^2 (T_g^{(h)} - T_r^{(h)})^2\right),
\eeq
where $Y_h$ and $\alpha_h$ are the Yukawa coupling and a fine
structure constant of the hidden gauge group, respectively, $N_y$
denotes a number of the final states for the process (b), $N_g$ is a
number of the generators of the gauge group, and $T_g^{(h)}$ and
$T_r^{(h)}$ are the Dynkin indices of the adjoint representation and the
matter fields in the representation $r$.  Although $x$ depends on the
structure of the SUSY breaking sector, its typical magnitude is
$O(10^{-2} - 10)$ for $m_\phi > \Lambda$.\footnote{
Roughly, we expect $N_{3/2} = O(1-10^2)$, $N_g = O(1)$, $\alpha_h =
0.1-1$, and $T_g^{(h)} - T_r^{(h)} = O(1)$, while $Y_h$ strongly
depends on the SUSY breaking models.
}
To be concrete we adopt $x=1$ as a reference value in the following, but it should be
kept in mind that there are  uncertainties in the gravitino
production rate in this case.

The abundance of gravitinos from the inflaton decay is therefore
\bea
Y_{3/2} &=& \frac{\Gamma_{3/2}}{\Gamma_\phi} \frac{3}{4} \frac{T_R}{m_\phi} \\
&\simeq& 1 \times 10^{-7} \,x \lrfp{g_*(T_{R})}{200}{-\frac{1}{2}} \lrfp{\la \phi \ra}{\GEV{15}}{2}
\lrfp{m_\phi}{\GEV{15}}{2} \lrfp{T_R}{\GEV{9}}{-1}.
\eea
Thus, the gravitino abundance increases as the inflaton VEV and mass.
Also, it should be noted that the gravitino abundance is inversely proportional to the
reheating temperature.

\begin{figure}[t]
\begin{center}
\begin{minipage}{16.4cm}
\centerline{
{\hspace*{0cm}\epsfig{figure=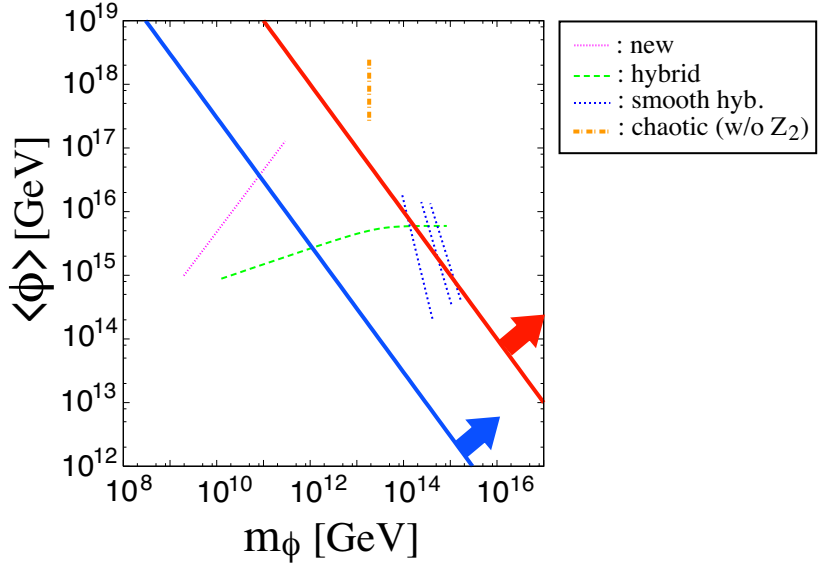,angle=0,width=11cm}}
}
\caption{
The gravitino-rich Universe is realized in the region above the blue solid line, while
the gravitino dominates the Universe in the region above the red solid line.
We have set $x=1$, $T_R = \GEV{9}$ and $m_{3/2} = \TEV{3}$.
}
\label{fig:inf}
\end{minipage}
\end{center}
\end{figure}

The Universe will be gravitino-rich, if
\beq
\la \phi \ra > 3 \times \GEV{12} \,\, x^{-\frac{1}{2}}
\lrfp{g_*(T_{R})}{200}{\frac{1}{4}}
\lrfp{m_\phi}{\GEV{15}}{-1} \lrfp{T_R}{\GEV{9}}{\frac{1}{2}},
\label{inf1}
\eeq
and the gravitino dominates the Universe if
\bea
&& \la \phi \ra > 1 \times \GEV{15} \,\, x^{-\frac{1}{2}}
\lrfp{g_*(T_{3/2})}{80}{-\frac{1}{8}} \lrfp{g_*(T_{R})}{200}{\frac{1}{4}}
\lrfp{m_{3/2}}{\TEV{3}}{\frac{1}{4}} \non\\
&&\qquad\,\, \times \lrfp{m_\phi}{\GEV{15}}{-1} \lrfp{T_R}{\GEV{9}}{\frac{1}{2}}.
\label{inf2}
\eea
The regions satisfying (\ref{inf1}) or (\ref{inf2}) are shown in Fig.~\ref{fig:inf},
together with representative inflation models, new~\cite{Asaka:1999jb,Nakayama:2011ri},
hybrid~\cite{Copeland:1994vg}, smooth hybrid~\cite{Lazarides:1995vr}
and chaotic~\cite{Kawasaki:2000yn} inflation. (The effect of the constant term in the superpotential on the
inflaton dynamics~\cite{Buchmuller:2000zm, Nakayama:2010xf} is not taken into account in this figure.)
We have set $x=1$, $T_R = \GEV{9}$ and $m_{3/2} = \TEV{3}$.
One can see that the Universe can be gravitino-rich for a large portion of the parameter
space, and this region will be larger for lower $T_R$.
For instance, in the smooth hybrid inflation, the inflaton mass and VEV are
of order $\GEV{15}$. In the chaotic inflation without a $Z_2$ symmetry,
the inflaton mass is about $2 \times \GEV{13}$, and the VEV is around the
Planck scale. For such high-scale inflation models, the resultant gravitino abundance
is so large that the gravitino dominated Universe can be realized.

So far, we have assumed the absence of the SUSY breaking singlet field $z$.
When it is present, it generally enhances the gravitino production rate through
the operators like $K \supset |\phi|^2 z$, or $|\phi|^2 zz$. The former induces a kinetic
mixing between the inflaton and $z$, while the latter induces the pair goldstino
production. Furthermore, such a singlet $z$ may be produced as coherent oscillations,
and it decays into gravitinos. Thus, the gravitino-rich Universe will be more plausible
in the presence of such a singlet.

\section{Cosmology}
\label{sec:4}
In this section we discuss cosmology of the gravitino-rich Universe.
In particular we study the neutralino DM produced by the gravitino decay,
and estimate the baryon asymmetry in thermal leptogenesis in the presence
of entropy dilution.

\subsection{Neutralino dark matter abundance}

We assume that the lightest neutralino $\chi$ is the LSP.
It decouples from thermal bath at the freeze-out
temperature, $T_f\sim m_{\chi}/28$ for the neutralino mass $m_\chi$
around or above a few hundred GeV.
Since $T_{3/2}$ is lower than $T_f$ for the gravitino at the PeV scale,
the neutralino abundance produced by gravitino decay is determined
according to the Boltzmann equation,
\bea
\frac{d n_\chi}{d t} + 3 H n_\chi = -\langle \sigma_\chi v_{\rm rel} \rangle
n^2_\chi,
\eea
where $n_\chi$ is the number density of $\chi$.
The effective annihilation cross section of the lightest neutralino is
parameterized by
\bea
\langle \sigma_\chi v_{\rm rel} \rangle=\frac{c_\chi}{m^2_\chi}.
\eea
One then finds the relic abundance of $\chi$ to be \cite{Fujii:2001xp,ArkaniHamed:2004yi}
\bea
\left(\frac{n_\chi}{s}\right)^{-1} =
\left(\frac{n_\chi}{s}\right)^{-1} \Big|_{T=T_{3/2}}
+\left(\frac{H}{s \langle \sigma_\chi v_{\rm rel} \rangle}\right)^{-1} \Big|_{T=T_{3/2}}.
\eea
When the gravitino abundance is large enough, the annihilation among the neutralinos
will take place until the expansion rate $H$ becomes equal to the annihilation rate.
This results in
\bea
\frac{n_\chi}{s}
\simeq
2.1\times 10^{-12}
\left(\frac{g_\ast(T_{3/2})}{80}\right)^{-1/4}
\left(\frac{c_\chi}{10^{-2}}\right)^{-1}
\left(\frac{m_\chi}{300{\rm GeV}}\right)^{2}
\left(\frac{m_{3/2}}{10^3 {\rm TeV}}\right)^{-3/2},
\eea
for $Y_{3/2}\gtrsim 10^{-12}$.

\begin{figure}[t]
\begin{center}
\begin{minipage}{16.4cm}
\centerline{
{\hspace*{0cm}\epsfig{figure=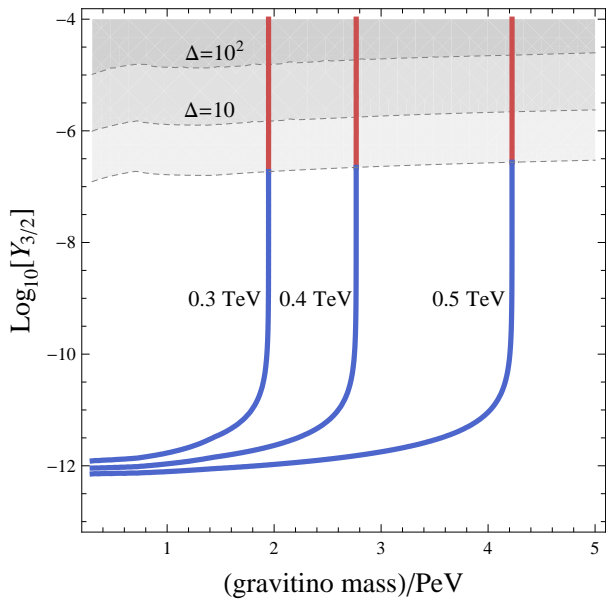,angle=0,width=8cm}}
}
\caption{
Wino DM density from gravitino decay consistent with
the observation, for $m_\chi=0.3$, 0.4, 0.5 TeV, respectively.
The left side of the solid line for given $m_\chi$ is excluded
by the overclosure constraint.
The gravitino dominates the Universe in the shaded region.
}
\label{fig:DM}
\end{minipage}
\end{center}
\end{figure}

In the gravitino-rich Universe, the lightest neutralino needs to have a large
annihilation cross section in order not to overclose the Universe.
This implies that, even if there are neutralinos thermally produced before the gravitino
decay, their thermal relic density will be small.
Let us consider the case where $\chi$ is dominated by the neutral
Wino.\footnote{
See Ref.~\cite{Jeong:2011sg} for the higgsino DM in the heavy gravitino
scenario, and also Ref.~\cite{Ibe:2011aa, Buchmuller:2012bt} for the neutralino DM
from gravitino decay in the case with $Y_{3/2}$ less than $10^{-12}$.
}
The Wino-like neutralino has $c_\chi$ around $10^{-2}$,
where we include the coannihilation effect but it is not so large because $T_{3/2}$
is similar in size to the mass difference of the charged and neutral Winos.
Fig. \ref{fig:DM} shows the parameter region where the Wino-like neutralinos
from gravitino decay account for the observed DM density.
For $Y_{3/2}\gtrsim 10^{-12}$, the neutral Wino with mass of several hundred
GeV can constitute the DM of the Universe.
We also plot the contours of the entropy dilution factor $\Delta$ for the
gravitino-domination case, $Y_{3/2}\gtrsim 10^{-7}$.
One can see from the figure that the Wino DM density is independent of the
gravitino abundance in the gravitino-rich Universe, and that the correct DM
abundance can be explained for the Wino-like neutralino with mass of several hundred GeV
and the gravitino at the PeV scale. In the next section we will discuss how to realize
such hierarchy between the gaugino mass and the gravitino mass.

Meanwhile, since the charged Wino is nearly degenerated in mass with the neutral
Wino, Sommerfeld effect should be included when the Wino has mass around
$4\pi m_W/g^2 \sim 2$ TeV \cite{Hisano:2003ec}.
However the resulting annihilation cross section would not be sufficiently large
to avoid the overclosure constraint for the gravitino mass at the PeV scale.

\subsection{Baryon asymmetry in leptogenesis}

One plausible way to generate the baryon asymmetry is through leptogenesis
\cite{Fukugita:1986hr},
and the simplest realization is to use lepton number violation of
the right-handed neutrinos in the seesaw mechanism~\cite{seesaw}.
Here we consider thermal leptogenesis in the gravitino-rich Universe.

The observed baryon asymmetry of the Universe can be explained
 through thermal leptogenesis if the reheating temperature after
inflation is high enough, $T_R\gtrsim \GEV{9}$ \cite{Buchmuller:2005eh}.
On the other hand, the gravitino abundance increases in proportional
to $T_R$ when produced by thermal scattering or pseudo-modulus decay.
Hence one may worry about the overproduction of DM by the gravitino decay
for the reheating temperature required by the successful thermal leptogenesis.
In order to avoid the tension,
$T_R$ must be chosen to be close to about $\GEV{9-10}$
for $m_{3/2} = 10 {\rm\,TeV} \hyp 100$\,TeV.\footnote{
Interestingly, the neutrino mass anarchy and the observed neutrino mass squared differences
suggest $T_R = \GEV{9}  \,{\rm \mathchar`-}\, \GEV{10}$,  when combined
with thermal leptogenesis.\cite{Jeong:2012zj}
}

Such a tension can be relaxed if the neutralino LSP has mass of several hundred GeV, while
the gravitino mass is at the PeV scale, as the neutralino DM abundance becomes
independent of the gravitino abundance.
In addition, if the gravitino dominates the Universe, there is an entropy dilution.
When the gravitinos are produced by thermal scatterings or the pseudo-modulus decay,
the entropy dilution factor is  approximately proportional to the reheating temperature,
implying that the present baryon asymmetry reads
\bea
\frac{n_B}{s} &\simeq& 3 \times 10^{-10} \Delta^{-1}  \lrf{\kappa}{0.1} \lrf{M_1}{\GEV{10}}
\lrf{m_{\nu 3}}{0.05{\rm \,eV}}
\delta_{\rm eff}
\propto \frac{M_{1}}{T_R},
\eea
where $\kappa$ denotes the efficiency factor,  $m_{\nu 3}$ the
heaviest neutrino mass, and $\delta_{\rm eff}$   the effective CP phase.
Here we consider the lepton asymmetry generated by the decay
of the lightest right-handed neutrino with mass $M_{1}(\lesssim T_R)$,
which is converted to the baryon asymmetry via the sphaleron effect.
Noting that the maximum value of the baryon asymmetry is realized for
$M_1 \sim T_R$, and that the dilution factor scales as $\Delta \propto T_R$,
one can see that the maximum value of baryon asymmetry
from leptogenesis will be independent of $T_R$.

Thus, if the gravitino-dominant Universe and the mass hierarchy between the LSP and the gravitino
are realized, both the maximum baryon asymmetry in thermal leptogenesis and the neutralino DM abundance
are independent of $T_R$. This feature has an interesting implication that thermal leptogenesis
can be implemented at high reheating temperatures while avoiding
overproduction of DM.

\begin{figure}[t]
\begin{center}
\begin{minipage}{16.4cm}
\centerline{
{\hspace*{0cm}\epsfig{figure=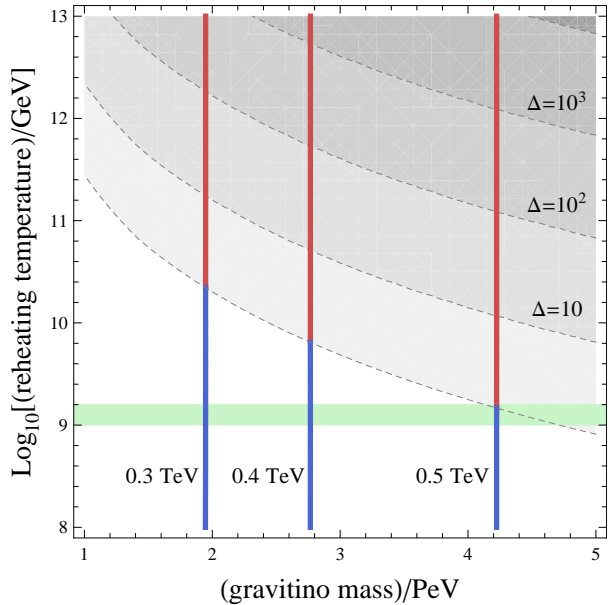,angle=0,width=8cm}}
}
\caption{
Gravitino-rich Universe in the pseudo-modulus scenario with
$m_z=4\times 10^8$ GeV.
The solid lines correspond to the contours of Wino DM
density from gravitino decay consistent with the observation,
for $m_\chi=0.3$, 0.4, 0.5 TeV, respectively.
The gravitino dominates the Universe in the shaded region.
For successful thermal leptogenesis, the reheating temperature after
inflation should be higher than about $10^9$ GeV. The maximum value of the
baryon asymmetry is constant along the (red) solid line in the shaded region.
}
\label{fig:Pseudo}
\end{minipage}
\end{center}
\end{figure}

Let us see the case where the gravitino produced by pseudo-modulus decay
dominates the Universe.
In Fig. \ref{fig:Pseudo}, we show the contours of Wino DM density
consistent with the observation in the ($m_{3/2},T_R$) plane,
for the pseudo-modulus with mass $m_z=4\times 10^8$ GeV.
Along the solid (red) lines in the shaded region, the gravitino dominates the Universe, and the maximum value
of the baryon asymmetry in thermal leptogenesis is constant because  $\Delta \propto T_R$.
For heavier (lighter) $m_z$, the pseudo-modulus abundance decreases, and the contour of the entropy
dilution factor will shift upwards (downwards).

\section{Mass hierarchy between gaugino and gravitino}
\label{sec:5}

We have seen in the previous section that the Wino-like neutralino LSP of mass several hundred GeV
produced by the decay of the gravitino at the PeV scale
can account for the observed DM  density.
In this section we discuss how to realize such mass hierarchy between the Wino-like neutralino
and the gravitino.

So far we have not specified how the SUSY breaking is mediated to the visible sector.
In the gravity mediation, both gaugino and scalar masses arise from coupling with a SUSY breaking singlet,
and all the superparticles have masses comparable to the gravitino mass. In this case,
the Wino mass can be much lighter than the other SUSY particles only at the price
of fine-tuning the coupling. Since the naturalness is not a serious issue in the high-scale SUSY,
we cannot exclude this possibility, and indeed such fine-tuning may be required by the anthropic
condition on the DM abundance.

In the absence of such SUSY breaking singlet,  on the other hand,
the gaugino mass can be hierarchically lighter than the gravitino mass.
In the heavy gravitino scenario, the anomaly mediation gives
dominant  contributions to the gaugino mass~\cite{Randall:1998uk,Giudice:1998xp}:
\bea
M_{\lambda} \;\sim\; \frac{m_{3/2}}{8\pi^2},
\label{m12}
\eea
unless one considers a specific SUSY breaking scenario such as no-scale type
models. Scalar masses are more model-dependent; the scalar fields obtain masses $\sim m_{3/2}$ through
contact interactions with a SUSY breaking field in the K\"ahler potential,
which are allowed even if the SUSY breaking field is charged under some
symmetry.  However, if  the visible sector is dynamically
sequestered from the SUSY breaking sector, it is possible that the scalar masses are
suppressed.
Compared to the other soft terms, therefore, the gaugino masses are less model-dependent,
and have a simple form \cite{Choi:2007ka}.

To arrange a light Wino of mass several hundred GeV for the gravitino mass at PeV scale,
the anomaly-mediation relation (\ref{m12}) is not sufficient,
and  one needs further suppression.\footnote{
In no-scale models, the chiral compensator auxiliary field is much suppressed
compared to $m_{3/2}$, and consequently anomaly-mediated soft masses are much
smaller than $m_{3/2}/8\pi^2$.
This provides another way to have relatively light gauginos in the heavy
gravitino scenario.
}
One way is to introduce SM vector-like matter fields $\Psi+\bar\Psi$ which
obtain masses from the VEV of a SM singlet so that their loops generate
soft masses via gauge mediation.
Let us consider $N_\Psi$ pairs of $\Psi+\bar\Psi$ forming ${\bf 5}+\bar{\bf 5}$
representation of SU(5) with
\bea
W = y_\Psi S \Psi\bar\Psi + f(S),
\eea
for which the gauge coupling unification of the MSSM is not spoiled.
Including the contribution from the messenger loops,
low energy gaugino masses read
\bea
\label{wino-mass}
\frac{M_a}{g^2_a} = \frac{b_a}{16\pi^2}m_{3/2}
-\frac{N_\Psi}{16\pi^2} \frac{F^S}{S},
\eea
where $(b_1,b_2,b_3)=(33/5,1,-3)$ are the MSSM beta function
coefficients.

The anomaly-mediated Wino mass can be cancelled when $S$ is stabilized in such a way
that $(F^S/S)/m_{3/2}$ is positive and order unity.
A simple way to achieve this is to consider a composite $S$ having an Affleck-Dine-Seiberg
superpotential \cite{Okada:2002mv},
\bea
f(S) = \frac{\Lambda^{3+n}_c}{S^n},
\eea
where $\Lambda_c$ is the dynamical scale much lower than the GUT scale, and $n$ is
a positive rational number.
Then the $F$-term scalar potential from the above superpotential competes with
the associated $A$-term to fix $S$ at $(\Lambda_c/m_{3/2})^{1/(n+2)}\Lambda_c$ with
\bea
\frac{F^S}{S} = \frac{2}{n+1}\Big(1+{\cal O}\Big(\frac{m^2_S}{m^2_{3/2}}\Big)
\Big) m_{3/2},
\eea
where $m^2_S$ is the soft mass squared of $S$.
Assuming that $S$ acquires soft mass around $m_{3/2}/8\pi^2$, one finds
that the Bino and Gluino have masses around $m_{3/2}/8\pi^2$ whereas
the Wino obtains a relatively light mass for $n=2N_\Psi-1$:
\bea
M_{\tilde W} \sim \frac{m_{3/2}}{(8\pi^2)^2},
\eea
which is a few hundred GeV for the gravitino at the PeV scale.
Since $S$ is a composite, the messengers have a small Yukawa coupling
$y_\Psi \sim \Lambda_c/M_P$, and get a SUSY preserving mass
$M_\Psi \sim (\Lambda_c/m_{3/2})^{1/(n+2)}\Lambda^2_c/M_P$.
Thus a dynamical scale higher than about $10^{12}$ GeV leads to
$M_\Psi\gg m_{3/2}$ as required for the relation (\ref{wino-mass}) to be valid.

Alternatively, in order to fix $S$ at a vacuum giving a positive $(F^S/S)/m_{3/2}$,
one may consider the model with
\bea
K &=& |Z|^2 +\sum_{\Phi} \left( |\Phi|^2  +
\frac{\kappa_\Phi}{3} \frac{|Z|^2}{M^2_P} |\Phi|^2 \right),
\nonumber \\
f(S) &=& y \Sigma (M^2 - S S^\prime),
\eea
for $\Phi=\{\Sigma,S,S^\prime,\Psi,\bar\Psi \}$ and $yM\gg m_{3/2}$,
so that $S$ is stabilized near the $F$-flat direction, $SS^\prime=M^2$.
Here $Z$ is the SUSY breaking field with $F$-term $\sim m_{3/2}M_P$,
and $\kappa_\Phi=1$ if $\Phi$ resides in a sector sequestered
from the SUSY breaking sector.
For $y_\Psi$ of order unity, the ratio between the soft scalar masses of $S$ and
$S^\prime$ is naively estimated by
\bea
\frac{m^2_{S^\prime}}{m^2_S} \sim
\frac{(8\pi^2)^2(1-\kappa_{S^\prime}) + y^2}{(8\pi^2)^2(1-\kappa_{S}) + y^2_\Psi}.
\eea
Let us assume $m^2_S\gg m^2_{S^\prime}$, which would be the case when $\kappa_S$
is less than one while $\kappa_\Phi=1$ for the others, or when $\kappa_\Phi=1$
for all the involved fields and $y\ll y_\Psi$.
Then the scalar potential develops a minimum at
$|S^\prime|^2 \simeq (m^2_S/m^2_{S^\prime})M^2 \gg |S|^2$ with
\bea
\frac{F^S}{S} =\left(1+{\cal O}\Big(\frac{m^2_{S^\prime}}{m^2_S} \Big) \right) m_{3/2},
\eea
and therefore the anomaly-mediated Wino mass is cancelled for $N_\Psi=1$,
making it possible to have $M_{\tilde W} \sim m_{3/2}/(8\pi^2)^2$ for $m^2_S$ larger
than about $8\pi^2 m^2_{S^\prime}$.
Meanwhile, the messenger mass reads $M_\Psi\sim (m^2_{S^\prime}/m^2_S)^{1/2} M$.

It is interesting to note that the above model possesses a global U(1),
under which $S$ and $S^\prime$ have charges of opposite sign while $\Sigma$ is neutral.
This can be identified as the Peccei-Quinn (PQ) symmetry solving the strong CP problem
\cite{pq,strongcp}, with the axion scale fixed at $F_a \sim |S^\prime|$.
The axion also constitutes the DM with $\Omega_a \sim 0.4 \theta^2_i(F_a/10^{12}\rm GeV)^{1.18}$
where $\theta_i$ is the initial misalignment.
The axion relic energy density will be diluted if the gravitino decays after
the QCD phase transition, for which $F_a$ larger than $10^{12}$ GeV
is allowed even for large initial misalignment.

Let us move to the possibility of arranging a light Wino through the higgsino
loops.
Including higgsino threshold effects, the gaugino masses are written
\cite{Giudice:1998xp,Gherghetta:1999sw}
\bea
\frac{M_a}{g^2_a} = \frac{b_a}{16\pi^2} m_{3/2}
+ \alpha \frac{k_a}{16\pi^2}m_{3/2},
\eea
for $(k_1,k_2,k_3)=(3/5,1,0)$, and $\alpha$ defined by
\bea
\alpha = \frac{\mu\sin2\beta}{m_{3/2}}
\frac{m^2_A}{|\mu|^2-m^2_A}
\ln\left(\frac{|\mu|^2}{m^2_A}\right),
\eea
where $\mu$ is the higgsino mixing parameter, $m_A$ is the mass of the heavy Higgs
bosons, and $\tan\beta$ is the ratio of the Higgs VEVs.
For a large $\mu$ around $m_{3/2}$ and low $\tan\beta$, the higgsino correction
can cancel the anomaly-mediated Wino mass, leading to a light Wino
at the price of fine-tuning.

Finally we briefly discuss phenomenological aspects of the model, where the lightest
neutralino $\chi$ and the lightest chargino $\chi^\pm$ are both Wino-like.
For large $\mu$, the dominant mass splitting between $\chi$ and $\chi^\pm$ comes
from gauge boson loops, and makes the chargino heavier than the neutralino
by about 160 MeV for $M_{\tilde W}$ around several hundred GeV \cite{Feng:1999fu}.
Hence $\chi^\pm$ dominantly decays into $\chi$ plus soft $\pi^\pm$ with a decay length
of a few {\it cm}, or longer if boosted, while producing a visible track in the detector.
Meanwhile, there are cosmological and astrophysical constraints arising since
the neutral Wino has a large annihilation cross section into a W-boson pair,
which require the Wino mass to be larger than about 300 GeV \cite{Ibe:2012hu}.

\section{Discussion and Conclusions}
\label{sec:6}

Being the superpartner of the graviton, the gravitino is coupled to all the sectors
with Planck-suppressed interactions. Thus, once the gravitino dominates the Universe,
it decays into all the lighter degrees of freedom, including those in the visible sector.
Then the question is how to generate the baryon asymmetry and DM.
In Sec.\ref{sec:4} we have considered thermal leptogenesis and the Wino-like
neutralino LSP non-thermally produced by the gravitino decay.  Here we briefly discuss
other possibilities.

One is the so-called gravitino-induced baryogenesis \cite{Cline:1990bw}.
In the presence of an R-parity and
baryon-number violating operator, $U_i D_j D_k$, and the associated $A$-term with a CP
phase, the gravitino decay can generate a right amount of the baryon asymmetry.
Since the coefficient of the R-parity violating operator must be sizable for the mechanism
to work,  any pre-existing baryon asymmetry would be washed out in this scenario.
Also, we need a source of CP violation; if there is a SUSY breaking singlet field,
it can satisfy the requirement. The presence of such SUSY breaking singlet generally
enhances the gravitino production from the inflaton decay, and furthermore, its
coherent
oscillations can decay into gravitinos.
Thus, the presence of the SUSY breaking singlet makes both the gravitino dominance
and the gravitino-induced baryogenesis  plausible.

Another one is the Affleck-Dine mechanism \cite{Affleck:1984fy}. The AD
mechanism is so efficient that a sufficient amount of baryon asymmetry can be generated
even in the presence of entropy production by the gravitino decay.
The AD mechanism in high-scale SUSY was explored in detail in Ref.~\cite{Higaki:2012ba}.

One of the plausible candidate for DM is the QCD axion. If the gravitino
decays after the QCD phase transition, the QCD axion abundance is diluted, allowing
a larger value of the axion decay constant~\cite{Kawasaki:1995vt}.

So far we have assumed that the gravitino is heavier than all the MSSM particles.
However, it is also possible that the gravitino decay into some of the MSSM
particles is kinematically forbidden. In this case the gravitino lifetime becomes
longer, and the gravitino mass is required to be heavier in order to avoid the tight BBN bound.

If there are additional light degrees of freedom such as the axion and axino, or
any light degrees of freedom in the hidden sector, the gravitino will decay into
those degrees of freedom.
The branching fraction into these hidden particles can be sizable especially
if some of the MSSM particles are heavier than the gravitino.
Interestingly, there is an argument that the string theory contains a plenitude
of string axions~\cite{Arvanitaki:2009fg}.
If the gravitino
decays into those axions and axinos with a sizable branching fraction, the produced
axions (and axinos) can account for the dark radiation hinted by the recent
observations~\cite{Ichikawa:2007jv}.
Also, if some of the hidden sector particles thus produced are stable in a cosmological
time scale, they will contribute to the DM; depending on its mass it can be
hot or warm DM.

In this paper, we have studied the gravitino cosmology in high-scale SUSY as suggested
by the recently discovered Higgs boson with mass $125 \hyp 126$\,GeV.
We have discussed various gravitino production processes
such as thermal production as well as
non-thermal production by the (pseudo) modulus and inflaton decay, in order to see if
the gravitinos are abundantly produced in the early Universe.  We have shown that
the Universe can be gravitino-rich for a wide range of parameters, and it is even possible
for the gravitino to dominate the Universe.  In the gravitino-rich Universe, the neutralino
LSP must have a large annihilation cross section, since otherwise the LSPs produced
by the gravitino decay would  overclose the Universe. We focused on the Wino-like LSP
and estimated its abundance as a function of the gravitino mass and abundance.
We also discussed how to realize the required mass hierarchy between the lightest neutralino
with $m_\chi = {\cal O}(100)$\,GeV and the gravitino at the PeV scale.

Among various possibilities, we have found
that the pseudo-modulus in the SUSY breaking sector is a plausible and interesting
source of gravitinos.
The dilution factor by the gravitino decay is proportional to the reheating
temperature in this case, which makes the maximum value of the baryon asymmetry
in thermal leptogenesis independent of the reheating temperature.

If high-scale SUSY is realized in nature, perhaps the naturalness argument is not the right
guiding principle to understand the observable parameters in our Universe.
The SUSY breaking scale suggested by the SM-like Higgs boson mass ranges from $10$\,TeV up to PeV
or even higher, but we do not know the reason why it takes such a value.
It may be due to a competition between the bias in the landscape and the anthropic condition
such as inflation~\cite{Nakayama:2011ri} or entropy dilution by the modulus decay~\cite{Takahashi:2011as}.

The gravitino, being the superpartner of the graviton, is long-lived and has universal Planck-suppressed couplings.
There are various processes in which gravitinos are produced.
Therefore the gravitino likely plays an important role in cosmology, especially
if high-scale SUSY is realized.
Our Universe might have been gravitino-rich.

\section*{Acknowledgment}
This work was supported by the Grant-in-Aid for Scientific Research on Innovative
Areas (No.24111702, No. 21111006, and No.23104008) [FT], Scientific Research (A)
(No. 22244030 and No.21244033 [FT]),
and JSPS Grant-in-Aid for Young Scientists (B) (No. 24740135) [FT].
This work was also supported by World Premier International Center Initiative (WPI
Program), MEXT, Japan, and by Grants-in-Aid for Scientific Research from the Ministry
of Education, Science, Sports, and Culture (MEXT), Japan (No. 23104008 and No.
23540283 [KSJ]).

\end{document}